\documentclass[aps,pra,showpacs,twocolumn,amsmath,amssymb,groupedaddress]{revtex4-1}
\usepackage{graphicx}
\begin{document}

\title{Defect solitons supported by nonlocal PT symmetric superlattices}

\author{Sumei Hu$^{1,2}$,  Daquan Lu$^1$, Xuekai Ma$^1$, Qi Guo$^1$, and  Wei Hu$^{1,*}$}
\affiliation{$^1$Laboratory of Nanophotonic Functional Materials and Devices,
South China Normal University,  Guangzhou  510631, P. R. China}
\affiliation{$^2$Department of Physics, Guangdong University of
Petrochemical Technology,  Maoming 525000, P. R. China}
\affiliation{$^*$Corresponding author: huwei@scnu.edu.cn}

\begin{abstract}
The existence and stability of defect solitons supported
by parity-time (PT) symmetric superlattices with nonlocal
nonlinearity are investigated.  In the semi-infinite gap, in-phase
solitons are found to exist stably for positive or zero
defects, but can not exist  in the presence of negative defects with strong
nonlocality. In the first gap,  out-of-phase solitons are stable for positive or zero defects,
whereas in-phase solitons are stable for negative defects. The
dependence of soliton stabilities on modulation depth of the PT
potentials is studied. It is interesting that solitons can exist
stably for positive and zero defects when the PT potentials are
above the phase transition  points.
\end{abstract}

\maketitle 

\section{Introduction}
Over the last few years, people have paid much
 {attention} to parity-time (PT)
symmetric systems
\cite{PT2008-PRL030402,PT2010-PRA,PT2010-Bendix,PRA040101-2011,PT2010-PRL,PT2009-prl093902,nature-phys2010,PT2011-Abdullaev,PT2011-H.G.Li}.
Bender {\it et al} have
shown that a family of  Hamiltonian with PT symmetry can still
exhibit entirely real spectra,  {despite the non-Hermitian Hamiltonian} \cite{Bender-1999}.  PT symmetry
requires that the complex potential satisfies $V(x)=V^*(-x)$. For a
PT symmetric Hamiltonian, there exists a critical threshold above
which its eigenvalues are not real but become complex, and the
system undergoes a phase transition because of the spontaneous
PT symmetry breaking \cite{Bender-1999}. Quite recently the notion
of PT symmetry was introduced within the framework of optics
\cite{PT2008-PRL030402}. This suggestion was based on judicious
designs that involve both optical gain/loss regions and the process
of index guiding\cite{PT2008-PRL030402,PT2009-prl093902,PT2010-Zheng}.
Unusual properties such as unidirectional invisibility
\cite{PT2011-Zin} and nonlinear switching structures
\cite{PT2010-Ramezani,PT2010-Sukhorukov} have been found in PT symmetric
structures. PT symmetry breaking in synthetic optical systems has
been observed  {experimentally} in
semiconductor heterostructure \cite{PT2009-prl093902}, photorefractive crystals\cite{nature-phys2010},  {and {\it LRC} circuits}
\cite{PRA040101-2011}.

Soliton phenomena in optical lattices and superlattices have
attracted considerable
attention\cite{DS2007-OE,DS1997-prb,DS1996-PRB,DS2006-PLA,DS2005-PRL,DS2009076802-PRL,DS20112680-ol}.
Optical superlattices can be considered as regular lattices hosting a
periodic chain of defects. Superlattices are fascinating because the
structures exhibit collective properties not shared by either
constituent, and these properties can be controlled through
variation of the structural parameters \cite{DS1987-PRL}. Defect
solitons in PT symmetric local lattices and superlattices have been
studied and stable solitons are not found in the first gap for
positive or zero defects \cite{DPT2011-OE,Lu2011-OE}. However,
defect solitons in PT symmetric superlattices with nonlocal
nonlinearity have not been studied. It is noteworthy that the
nonlinearity in photorefractive media, in which R\"{u}ter {\it
et al} have observed the PT symmetric wave propagation
\cite{nature-phys2010}, is nonlocal nonlinearity
\cite{Segev1992-prl,Segev1994-prl,Segev1998-prl}. Nonlocality plays
an important role in many areas of nonlinear physics
\cite{Science1997-Snyder}, drastically
 {modifies} the properties of solitons
and  {improves} the soliton stability
\cite{Science1997-Snyder,prl2005-Xu,Buccoliero2007-PRL}.  Therefore
it is worthy to study the properties of solitons in PT symmetric
optical superlattices with nonlocal nonlinearity.

In this paper, we study defect solitons supported by the PT
symmetric superlattices in nonlocal nonlinear media. We find that the
nonlocality expands stability  ranges of solitons, especially in the
first gap. Both  {out-of-phase} and
in-phase solitons can be stable in the first gap for
 positive, zero, and negative defects.
In-phase solitons can exist stably in the semi-infinite gap for
positive or zero defects,  but exist unstably for negative defects.
When the modulation depth of the PT potentials is small, solitons can
exist stably for positive and zero defects, even if the PT
potentials are above the phase transition
 {points}.

\section{Model}
We consider the propagation of light beam in PT symmetric defective
superlattices with  Kerr-type  nonlocal nonlinearity. The evolution
of complex amplitude $U$ of the light fields can be described by
following dimensionless nonlinear Schr\"{o}dinger equation,
\begin{equation}\label{solution}
i\frac{\partial U}{\partial z}+\frac{\partial^2 U}{\partial x^2}
 +[V(x)+i W(x)]U+nU=0,
\end{equation}
\begin{equation}\label{nonlinear}
n-d\frac{\partial^2 n}{\partial x^2}=|U|^2,
\end{equation}
where $x$ and $z$ are the transverse and longitudinal coordinates,
respectively, $n$ is the nonlinear refractive-index
change, $d$ stands for the degree of nonlocality of the nonlinear
response.
This type of nonlinear response with a finite region of nonlocality exists in many real physical systems, for instance, the photorefractive crystals  used in the experimental observation of PT symmetry in optics  \cite{nature-phys2010}.
 When $d\rightarrow 0$, Eq. (\ref{nonlinear}) reduces to
$n=|U|^2$, and above equations  reduce to the local case.
$V(x)$ and $W(x)$ are the real and imaginary parts of
the PT symmetric defective superlattices, respectively, which
are assumed in this paper as
\begin{eqnarray}\label{realpotentials}
 V(x)&=&V_0[\epsilon_1 \cos^{2}(x)+(1-\epsilon_1) \sin^{2}(2x)]\nonumber \\
 &\times & [1+\epsilon\exp(-x^8/128)], \\
 W(x)&=&W_0\sin(2x).
\end{eqnarray}
Here $\epsilon$ and $\epsilon_1$ represent the strength of the
defect and the modulation parameter of superlattice,
respectively. When $0.1 \leq \epsilon_1 \leq 0.7$, Eq.
(\ref{realpotentials}) has the superlattice shape \cite{DS2007-OE}.
Without losing of generality, we take $\epsilon_1=0.5$ throughout
the paper. The parameters $V_0$ and $W_0$ represent the modulation
depth of the real and imaginary parts of the PT symmetric potentials,
respectively.

\begin{figure}[htbp]
\includegraphics[width=8.3cm]{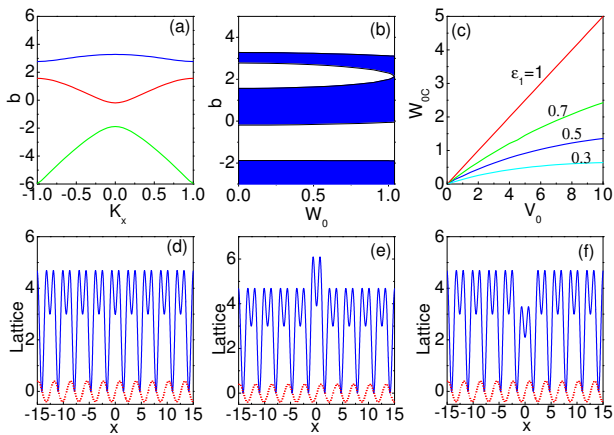}
   \caption{(Color online)  (a) Band structure of the superlattice with $V_0=6$, $W_0=0.4$ and $\epsilon_1=0.5$. (b) Band structure of the superlattice with different $W_0$ values, where $V_0=6$ and $\epsilon_1=0.5$. (c) The phase transition point $W_{0C}$ versus $V_0$ for different $\epsilon_1$ values.
(d)-(f) Lattice intensity profiles  {for the superlattice with defects}
   (d) $\epsilon=0.3$, (e) $\epsilon=0$, and (f)$\epsilon=-0.3$
   Blue solid lines and red dotted lines represent the real and imaginary parts,
   respectively.}  \label{Fig1}
\end{figure}

The PT symmetric superlattices in Eq. (\ref{solution}) have a Bloch band
structure when $\epsilon=0$ and $n=0$. The Bloch band structure
obtained by the plane wave expansion method for $\epsilon_1=0.5$,
$V_0=6$ and $W_0=0.4$ is shown in Fig.\ref{Fig1}(a). One can see
that the region of the semi-infinite gap is $b>3.27$, and the first
and second gaps are $1.56<b<2.77$ and $-1.88<b<-0.19$, respectively,
where $b$ is the propagation constant. Figure \ref{Fig1}(b) shows
the Bloch band structure for $V_0=6$ with different $W_0$ values.
The region of the first gap
decreases with increasing $W_0$. The first gap disappears at
$W_0=1.035$, which is the phase transition point $W_{0C}$. When the
system is above the phase transition point, i.e. $W_0>1.035$, the
band structure becomes complex. We find that, for the PT symmetric superlattices, $W_{0C}$ changes with $V_0$ nonlinearly, as shown in Fig. \ref{Fig1}(c). However, for the ordinary
lattices($\epsilon_1$=1), the $W_{0C}$ increases linearly with
increasing $V_0$, i.e, $W_{0c}/V_0=0.5$ \cite{PT2008-PRL030402,
PT2010-PRA}. In this paper, we study the PT symmetric superlattices {below or above the phase transition points}, respectively.
Figures \ref{Fig1}(d)-\ref{Fig1}(f) show the intensity distributions of the PT
superlattice potentials with the strength of defects $\epsilon=0$,
$\epsilon=0.3$ and $\epsilon=-0.3$, respectively. $\epsilon=0$
corresponds to the uniform superlattice.

We search for stationary solutions to Eqs. (\ref{solution}) and
(\ref{nonlinear}) in the form $U=f(x)\exp(ibz)$, where  $f(x)$ is
a complex function and satisfies equations,
\begin{equation}
 bf=\frac{\partial^2 f}{\partial x^2} +[V(x)+i W(x)]f+nf, \label{field}
\end{equation}
\begin{equation}
 n - d \frac{\partial^2 n}{\partial x^2}=|f|^2. \label{nonfield}
\end{equation}
The solutions of defect solitons are gotten numerically from Eqs. (\ref{field}) and  (\ref{nonfield})  {by the
modified squared-operator method\cite{yang-2007}} and  shown in the next
section.
To elucidate the stability of defect
solitons, we search for the perturbed solutions to Eqs.
(\ref{solution}) and (\ref{nonlinear}) in the form  {
$U(x,z)=[f(x)+u(x,z)+i v(x,z)] \exp(ibz)$}, where the real
[$u(x,z)$] and imaginary [$v(x,z)$] parts of the perturbation can
grow with a complex rate $\delta$ upon propagation,
 { i.e. $u(x,z)=u(x)\exp(i\delta z)$
and $v(x,z)=v(x)\exp(i\delta z)$. Substituting the perturbed
solution into Eq. (\ref{solution}) and linearization of it} around
the stationary solution $f(x)$ yields the eigenvalue problem
\begin{eqnarray} \label{solution7}
 \delta v =& \frac{\partial^2 u}{\partial x^2}+ (n-b) u +(V u-W v)\nonumber \\
& +Re[f(x)]\int_{-\infty}^{\infty}2 G(x-\xi)u(\xi)Re[f(\xi)]{\rm d}\xi \nonumber \\
&  +Re[f(x)] \int_{-\infty}^{\infty} 2 G(x-\xi)Im[f(\xi)] v(\xi)
{\rm d}\xi,
\\
 \delta u  = & -\frac{\partial^2 v}{\partial x^2} -(n-b) v  -(Wu+Vv)\nonumber \\
& - Im[f(x)] \int_{-\infty}^{\infty} 2G(x-\xi)Im[f(\xi)] v(\xi) {\rm d} \xi \nonumber \\
& -Im[f(x)]\int_{-\infty}^{\infty} 2G(x-\xi)Re[f(\xi)] u(\xi) {\rm
d}\xi.
\end{eqnarray}
Here $G(x)=[1/(2\sqrt{d})]\exp(-|x|/\sqrt{d})$ is the response
function of the nonlocal media. Above eigenvalue problem is solved
numerically  {by the original-operator iteration method}\cite{Yang32} to find the maximum value of $Re(\delta)$. If
$Re(\delta)>0$, solitons are unstable. Otherwise, they are stable.

\section{Numerical Results}

\begin{figure}[htbp]
\includegraphics[width=8.3cm]{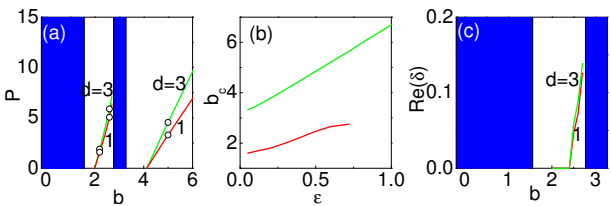}
    \caption{(Color online)  (a) Soliton power $P$ versus propagation constant $b$; (b) Cutoff point $b_c$ versus the depth of defects.
    (c) Unstable growth rate $Re(\delta)$ versus propagation constant $b$ for {out-of-phase} solitons in the first gap. For all cases $V_0=6$,  $W_0=0.4$, and $\epsilon=0.3$ except for (b).} \label{Fig2}
\end{figure}

In the nonlocal PT symmetric defective superlattices, we find two
types of defect solitons. The first type
is in-phase soliton, whose real part of optical fields is symmetric
and the imaginary part is antisymmetric. The other type is
 {out-of-phase} soliton, whose real
part is antisymmetric and the imaginary part is symmetric.  It is
noteworthy that if $U=f(x)\exp(ibz)$ is a solution to Eqs.
\ref{solution} and \ref{nonlinear}, we also have a series of
solution, i.e. $U=f(x)e^{i\theta}\exp(ibz)$, where $\theta$ is an
arbitrary initial phase. We know that  $f(x)$ and $f(x)e^{i\theta}$
represent the same physical soliton despite the initial phase,  {but their parities are uncertain}.
Due to PT symmetry, one can find a proper initial phase to  {guarantee that both the real and imaginary parts of solitons have fixed parity (odd or even)}.
When we use a real function as the initial trial in
the iteration program solving Eqs. (\ref{field}) and
(\ref{nonfield}), the final  {convergency} solution has a certain parity
and its real part is always bigger than its imaginary part. Then we
can distinguish the in-phase and  {out-of-phase} solitons in this occasion. We find that in the first gap,
 {out-of-phase} solitons can exist
stably for positive and zero defects, whereas in-phase solitons
can exist stably for  negative defects. In the semi-infinite gap, we
find in-phase solitons for both positive, zero, and negative
defects.

\begin{figure}
\includegraphics[width=8.3cm]{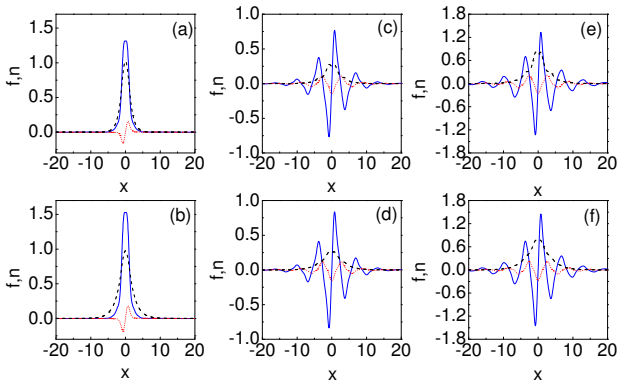}
    \caption{(Color online) The complex fields (solid blue: real part; dotted red: imaginary part) and refractive-index changes (dashed black) for in-phase solitons in the semi-infinite gap at (a) $b=5$, $d=1$, and (b) $b=5$, $d=3$, and for  {out-of-phase} solitons in the first gap at (c) $b=2.2$, $d=1$, (d) $b=2.2$, $d=3$, (e) $b=2.6$, $d=1$, (f) $b=2.6$, $d=3$, respectively. For all cases  $\epsilon=0.3$, $V_0=6$ and $W_0=0.4$.} \label{Fig3}
\end{figure}

\begin{figure}
\includegraphics[width=8.3cm]{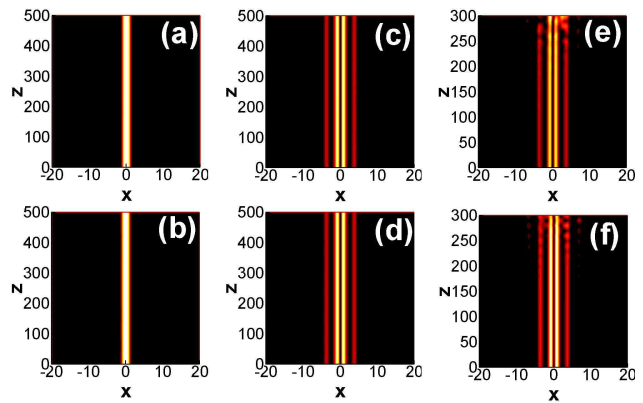}
  \caption{(Color online) (a)-(f)  {Evolutions} of defect solitons corresponding to Figs. \ref{Fig3}(a)- \ref{Fig3}(f), respectively.} \label{Fig4}
 \end{figure}

For positive defects, we assume $\epsilon=0.3$ and the results
are shown in Figs.~\ref{Fig2}-\ref{Fig4}. We find in-phase solitons in the semi-infinite gap and out-of-phase solitons in the first gap,  {respectively}.  Figure \ref{Fig2}(a) shows
that the power of solitons [defined as {$P=\int_{-\infty
}^{+\infty}|f(x)|^2 {\rm d}x$}] for both in-phase and
 {out-of-phase} solitons increases
almost linearly as increasing of propagation constant. There exists a
cutoff point of propagation constant, below which the defect
solitons vanish.  {The propagation constants of the cutoff points} do not depend on
the nonlocality degree. This feature is similar to the case of
traditional uniform lattices in nonlocal media
\cite{prl2005-Xu}. The propagation constants of the cutoff
 {points} for both
 {in-phase and out-of-phase} solitons increase with
increasing the depth of positive defects, as shown in Fig.
\ref{Fig2}(b). When $\epsilon>0.725$, the
 {out-of-phase} solitons can not exist
in the first gap.

We find that in-phase solitons are stable in the whole regime where
solitons exist in the semi-infinite gap when  {$W_0=0.4$}, like its counterparts in
local PT lattices.  {Out-of-phase}
solitons are found to exist stably for the low propagation constants
in the first gap, as shown in Fig. \ref{Fig2}(c). The stability
regions of defect solitons decrease with increasing  $W_0$. In-phase solitons in the semi-infinite gap  {become} unstable for large propagation when $W_0>0.46$ and their stability region  vanishes totally as $W_0>0.58$.  {Out-of-phase} solitons become unstable in the whole first gap when $W_0>0.53$. For comparison,
 {out-of-phase} solitons in the first
gap do not exist in local PT symmetric defective superlattices with
positive defects \cite{Lu2011-OE}. It shows that the nonlocality
expands the existing and stable ranges of solitons.

The field distributions of defect solitons are shown in Figs.
\ref{Fig3}(a)-\ref{Fig3}(f) for different nonlocal degrees  and
propagation constants, which correspond to the cases represented by circls in
Fig. \ref{Fig2}(a). Figures \ref{Fig3}(a) and \ref{Fig3}(b) show
that in-phase solitons exist in the semi-infinite gap, while solitons
in the first gap are  {out-of-phase}
as shown in Figs. \ref{Fig3}(c)-\ref{Fig3}(f). As the nonlocality
degree increases, soliton power increases but the shape of
their  {field distribution} changes very little. The propagations of
solitons are also simulated based on Eqs.~(\ref{solution}) and
(\ref{nonlinear}), and 1\% random-noise perturbations are added into
the initial input to verify the results of the linear stability
analysis. The propagations corresponding to solitons in Figs.
\ref{Fig3}(a)-\ref{Fig3}(f) are shown in  {Fig. \ref{Fig4}(a)-\ref{Fig4}(f)}. We can see
that in-phase solitons are stable in the semi-infinite gap and
 {out-of-phase} solitons are stable in
the first gap for the low propagation constants.

\begin{figure}
\includegraphics[width=5.6cm]{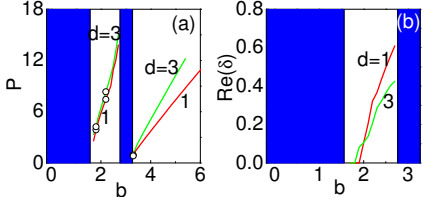}
    \caption{(Color online) (a) Soliton Power $P$ versus propagation constant $b$ for different value of $d$.
  (b) Unstable growth rate $Re(\delta)$ versus propagation constant $b$  for {out-of-phase} solitons in the first gap. For all  {cases} $V_0=6$, $W_0=0.4$, and $\epsilon=0.0$ }
\label{Fig5}
\end{figure}
\begin{figure}
\includegraphics[width=8.3cm]{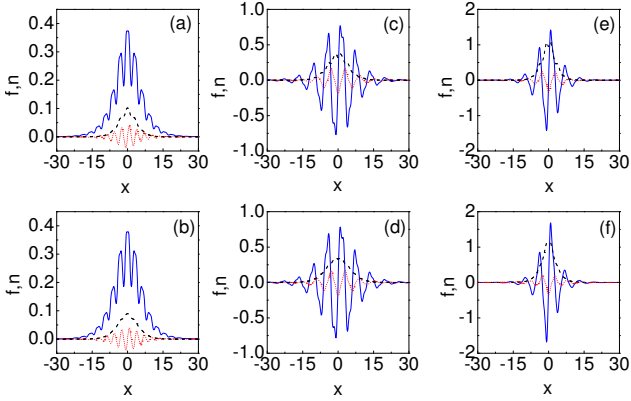}
  \caption{(Color online) The complex fields (solid blue: real part; dotted red: imaginary part) and refractive-index changes (dashed black) for in-phase solitons in
the semi-infinite gap at (a) $b=3.3$, $d=1$ and (b) $b=3.3$, $d=3$,
and for  {out-of-phase} solitons in the
first gap at (c) $b=1.8$, $d=1$, (d) $b=1.8$, $d=3$, (e) $b=2.2$,
$d=1$ and (f) $b=2.2$, $d=3$, respectively.} \label{Fig6}
\end{figure}
\begin{figure}
\includegraphics[width=8.3cm]{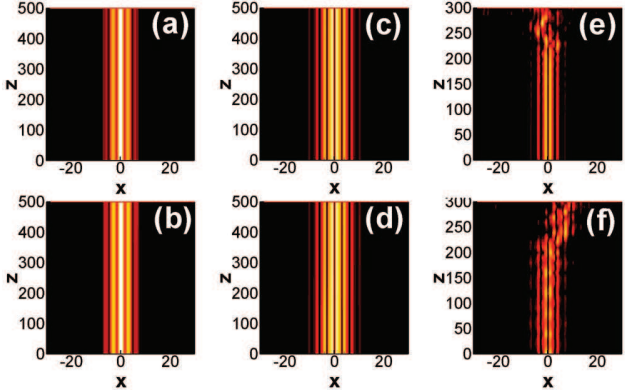}
  \caption{(Color online) (a)-(f): Evolutions of  defect solitons corresponding to Figs. \ref{Fig5}(c)- \ref{Fig5}(h), respectively.} \label {Fig7}
  \end{figure}

Figures \ref{Fig5}-\ref{Fig7} show the results for  { zero defects}
($\epsilon=0$). The properties of in-phase solitons
in the semi-infinite gap and out-of-phase solitons in the first gap for zero defects are almost the same as those for
 { positive defects}, except the positions of the cutoff points. From Fig.
\ref{Fig5}(a) we can see that the cutoff points  approach the edge
of the gap, so in-phase solitons  are stable in the whole
semi-infinite gap  {when
$W_0=0.4$}.
For comparison, the local defect solitons in the semi-infinite gap
are unstable near the edge of the gap\cite{Lu2011-OE}.  {For nonlocal superlattices, we find that in-phase solitons  are  always stable when $W_0<0.48$. As increasing $W_0$, the stable region begins to shrink, and in-phase solitons are unstable in the whole semi-infinite gap when $W_0>0.76$.}

Figure \ref{Fig5}(b) is the perturbation growth rate $Re(\delta)$
versus propagation constant for  {out-of-phase} solitons in the first gap with different nonlocality degree
when $W_0=0.4$. We can see that out-of-phase solitons are stable for the low propagation constant.  {As increasing $W_0$}, the stability region of  {out-of-phase} solitons vanishes after $W_0>0.51$.

The examples of in-phase solitons in the semi-infinite gap and
 {out-of-phase} solitons in the first
gap  {for zero defects} are shown in Figs.
\ref{Fig6}(a)- {\ref{Fig6}}(f), and
the corresponding propagations are shown in Fig. \ref{Fig7}. We can
see that in-phase solitons in the semi-infinite gap have more peaks
than those for positive defect. Figure \ref{Fig7} shows that
in-phase solitons are stable at the edge of the gap whereas
 {out-of-phase} solitons are stable with
low propagation constants. However,
 {out-of-phase} solitons in local PT
symmetric defective superlattices do not exist in the first gap for
zero defects \cite{Lu2011-OE}.

\begin{figure}
\includegraphics[width=8.3cm]{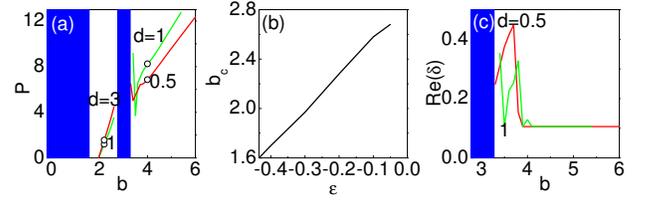}
    \caption{(Color online) (a) Soliton Power $P$ vs propagation constant $b$ for in-phase solitons  in the semi-infinite gap and  the first gap. (b)Cutoff point $b_c$ versus the depth of defects.  (c) Unstable growth rate $Re(\delta)$ versus propagation constant $b$ for the in-phase solitons in the semi-infinite gap. For all  {cases} $V_0=6$, $W_0=0.4$, and $\epsilon=-0.3$ except for (b).} \label{Fig8}
\end{figure}

\begin{figure}
\includegraphics[width=5.6cm]{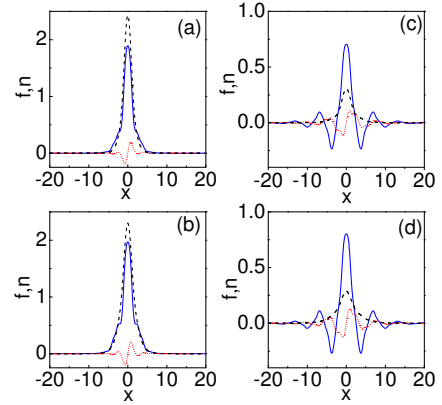}
    \caption{(Color online)
(a)-(d) The complex fields (solid blue: real part; dotted red: imaginary part)
and refractive-index changes (dashed black) for in-phase solitons in the
semi-infinite gap at(a) $b=4$, $d=0.5$ and (b) $b=4$, $d=1$, or in
the first gap at (c) $b=2.2$, $d=0.5$ and (d) $b=2.2$, $d=1$,
respectively.} \label{Fig9}
    \end{figure}

\begin{figure}
\includegraphics[width=5.6cm]{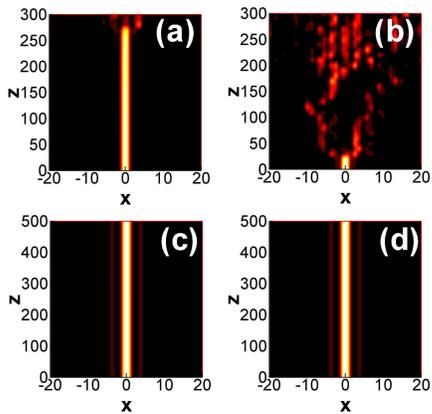}
  \caption{(Color online) (a)-(d): Evolutions of defect solitons corresponding to Figs. \ref{Fig8}(a)- \ref{Fig8}(d), respectively.} \label {Fig10}
  \end{figure}
For  {negative defects}, we assume $\epsilon=-0.3$ and the results are
shown in Figs. \ref{Fig8}- \ref{Fig10}. We find that in-phase
solitons in the semi-infinite gap exist only for a weak nonlocality
degree ($d<1.2$), and our numerical program can not find any
stationary solution for $d>1.2$.  {The linear stability analysis} shows
that in-phase solitons in the semi-infinite gap are unstable in
their whole existence regions,  {as shown} in Fig.~\ref{Fig8}(c).
In the first gap, differing from positive and zero defects, in-phase
solitons  {are found and stable} in their whole existence regions when $W_0<0.56$.
 {As increasing $W_0$, in-phase solitons in the first gap become
unstable} for the large propagation constant when
$W_0<0.78$. Solitons are all unstable after  $W_0>0.78$.

We can also see that there exists a cutoff point of propagation constant above which in-phase solitons exist. This
feature is similar to the case for positive and zero defects.  {Figure} \ref{Fig8}(b) shows that the cutoff point shifts toward the  {lower}
propagation constant with increasing the depth of negative defects,
and the existence region of in-phase solitons in the first gap
increases too. When $\epsilon=-0.435$,  {the cutoff point arrives at the edge of the gap and then in-phase solitons} exist stably in the whole first gap for $W_0=0.4$.

The examples of in-phase solitons in the semi-infinite gap and the
first gap are shown in Figs. \ref{Fig9}(a)-\ref{Fig9}(d), and the
corresponding propagations are shown in  {Figs. \ref{Fig10}(a)-\ref{Fig10}(d)}. One can see that in-phase solitons are unstable in the semi-infinite gap
whereas they are stable in the first gap.

\begin{figure}
\includegraphics[width=6.2cm]{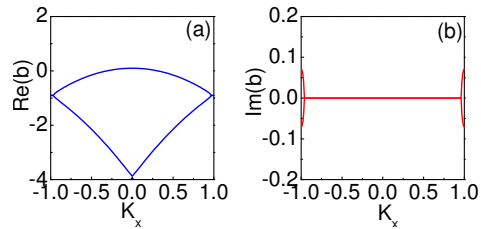}
  \caption{(a) Real and (b) imaginary parts of the bandgap structures for the superlattice with $\epsilon_1=0.5$, $V_0=0.2$ and $W_0=0.15$, respectively.} \label {Fig11}
  \end{figure}
\begin{figure}
\includegraphics[width=8.3cm]{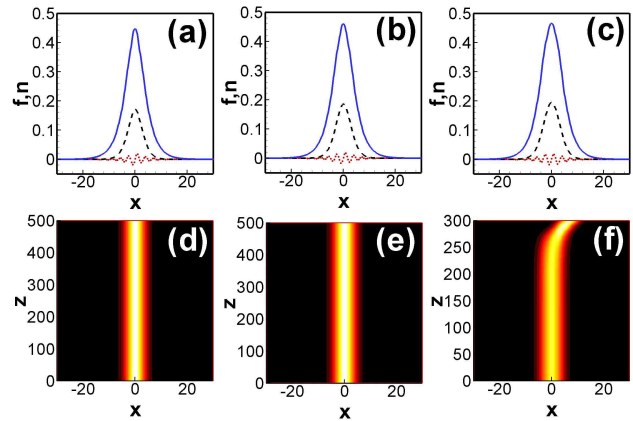}
    \caption{(Color online) The complex fields (solid blue: real part; dotted red: imaginary part) and refractive-index changes (dashed black) of solitons  for (a) positive defect $\epsilon=0.3$,  (b) zero defect $\epsilon=0$, and (c) negative defect $\epsilon=-0.3$. (d)-(f) Evolutions of defect solitons corresponding to these solitons
in (a)-(c), respectively. For all cases $V_0=0.2$, $W_0=0.15$,
$b=0.2$, and $d=1$.} \label{Fig12}
\end{figure}

Finally, we study the case of PT symmetric superlattices above the phase transition point. We find that  {in-phase solitons for positive and zero defects can exist stably even if PT potentials} are above the phase transition points when the value of $V_0$ is small.
We take $V_0=0.2$ and the phase transition point
$W_{0c}=0.05$, as shown in the fig. \ref{Fig1}(c). Figures
\ref{Fig11}(a) and \ref{Fig11}(b) show the real and imaginary parts
of bandgap structures for $V_0=0.2$ and $W_0=0.15$, respectively.
One can see that the corresponding band diagram is partially
complex. Figures. \ref{Fig12}(a)-\ref{Fig12}(c) show the examples of
solitons above the phase transition points  for positive ($\epsilon=0.3$), zero ($\epsilon=0$), and negative defects ($\epsilon=-0.3$),
respectively. Figures \ref{Fig12}(d)-\ref{Fig12}(f) show the
corresponding propagations of those solitons.  {Supported by the  nonlocal nonlinearity} ($d=1$), defect solitons for positive
and zero defects are stable when $W_0$ is above the phase transition
point, whereas  {they are}  unstable for negative defects. It is noted that the imaginary parts of the complex fields for solitons shown in Fig. \ref{Fig12} are still very small comparing with their real parts,  { although the PT potentials are above the  phase transition points.}

\section{Conclusion}

We have studied the existence and stability of defect
solitons supported by parity-time symmetric nonlocal
superlattices. Unusual properties are found in this system. The nonlocality can drastically affect the existence and stability of defect solitons. For  positive or zero defects,
in-phase solitons can exist stably in the semi-infinite gap
  and  {out-of-phase} solitons can exist stably in the first gap. For negative defects, {in-phase solitons  are  stable in the first gap, but unstable in the semi-infinite gap for the weak nonlocality degree}.
The values of $W_0$ and $V_0$ affect the soliton stability
strongly. When $V_0$ is large,  {defect solitons} are unstable  {unless}
$W_0$ is far below the phase transition point $W_{0C}$. However,  {when $V_0$ is small ,} the stable solitons can be found for positive and zero defects even if
$W_0$ is above the phase transition points. These properties of  defect
solitons in PT symmetric nonlocal superlattices are obviously
different from those in the local PT superlattices.

\acknowledgments
This research was supported by the National Natural Science Foundation of China (Grant Nos. 10804033, 11174090, and 11174091).

\end{document}